\documentclass[%
 aip,
rsi,%
 amsmath,amssymb,
preprint,%
]{revtex4-2}
\usepackage{hyperref}
\usepackage{graphicx}
\usepackage{dcolumn}
\usepackage{bm}
\usepackage{color}

\usepackage{subfig}
\usepackage{gensymb}

\begin{document}

\preprint{AIP/123-QED}

\title[Adhikari et al. 2023]{Effect of a guide field on the turbulence like properties of magnetic reconnection}

\author{S. Adhikari}
\email{subash.adhikari@mail.wvu.edu}

\affiliation{Department of Physics and Astronomy, University of Delaware, Newark, DE 19716, USA}

\affiliation{Department of Physics and Astronomy, West Virginia University, Morgantown, West Virginia 26506, USA}

\author{M. A. Shay}%

\affiliation{Department of Physics and Astronomy, University of Delaware, Newark, DE 19716, USA}

\affiliation{Bartol Research Institute, Department of Physics and Astronomy, University of Delaware, Newark, DE 19716, USA}

\author{T. N. Parashar}%
\affiliation{
School of Chemical and Physical Sciences,Victoria University of Wellington, Wellington 6012, NZ
}%
\affiliation{Department of Physics and Astronomy, University of Delaware, Newark, DE 19716, USA}

\author{W. H. Matthaeus}%
\affiliation{Department of Physics and Astronomy, University of Delaware, Newark, DE 19716, USA}

\affiliation{Bartol Research Institute, Department of Physics and Astronomy, University of Delaware, Newark, DE 19716, USA}

\author{P. S. Pyakurel} 
\affiliation{Space Sciences Laboratory, University of California, Berkeley, CA 94720, USA}%

\author{J.E. Stawarz} 
\affiliation{Department of Mathematics, Physics, and Electrical Engineering, Northumbria University, Newcastle upon Tyne NE1 8ST, UK}%

\author{J.P. Eastwood} 
\affiliation{Department of Physics, Imperial College London, SW7 2AZ, UK}%

\date{\today}

\begin{abstract}

The effect of an external guide field on the turbulence-like properties of magnetic reconnection is studied using five different $2.5$D kinetic particle-in-cell (PIC) simulations. The magnetic energy spectrum is found to exhibit a slope of approximately $-5/3$ in the inertial range, independent of the guide field. On the contrary, the electric field spectrum, in the inertial range steepens more with the guide field and approaches a slope of $-5/3$. In addition, spectral analysis of the different terms of the generalized Ohm's law is performed and found to be consistent with PIC simulations of turbulence and MMS observations. Finally, guide field effect on the energy transfer behavior is examined using von-K\'arm\'an Howarth (vKH) equation based on incompressible Hall-MHD. The general characteristics of  the vKH equation with constant rate of energy transfer in the inertial range, is consistent in all the simulations. This suggests that the qualitative behavior of energy spectrum, and energy transfer in reconnection is similar to that of turbulence,  indicating that reconnection fundamentally involves an energy cascade.

\end{abstract}

\maketitle

\section{\label{sec:introduction}Introduction}
Reconnection and turbulence are ubiquitous processes in plasmas, including laboratory, astrophysical, and space plasmas. Traditionally these processes have been studied independently of each other, except for a few studies scattered across the last few decades. These studies have associated magnetic reconnection with turbulence, focusing mostly on turbulent reconnection both theoretically~\cite{Strauss86,lazarian1999reconnection} and numerically \cite{matthaeus1986turbulent, smith2004hall, lapenta2008self,lazarian2015turbulent,lazarian20203d}. Other themes include examining reconnection in turbulence \cite{retino2007situ,servidio2009magnetic,zhou2020multi} or turbulence generation via reconnection related instabilities ~\cite{munoz2018kinetic,leonardis2013identification,PucciApJ2018,LapentaApJ2020}.

In a series of papers, we have been taking these connections further; 
we have probed a deep connection between reconnection and turbulence implied by morphological similarities between the two, including spectral features and energy transfer across scales \cite{adhikari2020reconnection, adhikari2021magnetic}. For example, the magnetic energy spectrum of antiparallel laminar reconnection exhibits a spectral slope of $-5/3$ in the inertial range~\cite{adhikari2020reconnection}. More importantly, a von K\'arm\'an Howarth analysis of the energy transfer across scales show that the cascade of energy is quite similar in both laminar reconnection simulations and traditional turbulence simulations~\cite{adhikari2021magnetic}. These results imply that there may exist a fundamental universality between magnetic reconnection and turbulence. Although the analyses so far have focused on antiparallel reconnection, magnetic reconnection occurs under very general conditions including lower magnetic shears as well as asymmetric situations.

Here we extend the previous studies to the case of guide field reconnection and determine how the degree of magnetic shear modifies the turbulence-like properties of reconnection. In this study, we simulate 2D reconnection for a range of guide fields and examine spectra and energy cascade properties in a similar fashion to previous work\cite{adhikari2020reconnection, adhikari2021magnetic}. At MHD scales, the guide field does not fundamentally modify the power-law magnetic spectrum nor the energy transfer/cascade. We also examine the power spectra of both the electric field and its constituent terms from Ohm's law.

The rest of the paper is structured as follows: In Section \ref{sec:theory} we provide a detailed explanation of the physical meaning of the von-K\'arm\'an Howarth equation (aka Third-Order law), while Section \ref{sec:simulations} introduces the types of reconnection simulations used in the analysis. In Section \ref{sec:results} we discuss the results and finally in Section \ref{sec:conclusions} we present our discussions and conclusions.

\section{\label{sec:theory}Physical meaning of the Von K\'arm\'an Howarth Equation}

The reconnection/turbulence connection continues to receive increasing scrutiny, so it is important to make key physical concepts accessible from one field to another. In particular, while the von K\'arm\'an Howarth analysis of turbulent cascades is a powerful tool used in the turbulence community, the reconnection community has had limited exposure to  with this formalism. In that light it is helpful to spend some effort to give some physical insight into the von K\'arm\'an Howarth equation. 

Outside the turbulence community, the impression of turbulence is completely dominated by the concept of spectral energy density and its $-5/3$ power law when plotted versus wavenumber $k$. However, the von K\'arm\'an Howarth third order law bypasses $k$ in favor of the concept ``lag,'' which represents a variation over a length $l$ and can intuitively be thought of as the inverse of the wavenumber $l \approx 1/k$. As with multidimensional spectral analysis, the lag is in general a vector $\boldsymbol{l}$. 

When discussing the properties of turbulent fluctuations $\mathbf{u}$ and $\mathbf{b}$, we will use Alfv\'enic units, where the fluctuation amplitude of magnetic field $\mathbf{b}$ has units of velocity $\mathbf{u}$, i.e., $\mathbf{b} \equiv (\mathbf{B} - \langle \mathbf{B} \rangle_r  )/\sqrt{4 \pi mn}$, where $\langle \dots \rangle_r$ is an average over regular space (not lag space). Note that we are assuming incompressible turbulence in this discussion so that for all intents and purposes the number density $n$ can be assumed to be a constant. Much of our discussion will involve functions of $\lvert\mathbf{u}\rvert^2$ and $\lvert\mathbf{b}\rvert^2$, which in Alfv\'enic units have units of energy per unit mass. Physically, the total energy of the system associated with these functions can be determined by multiplying by the mass density and integrating over the entire system in regular space. As a shorthand, we will write energy per unit mass as ``energy/mass'' in the manuscript.

In order to characterize fluctuations associated with a lag $\boldsymbol{l}$, the variation of turbulent fluctuations 
is defined as the ``increment,'' with the velocity increment $\delta \mathbf{u(r,}\, \boldsymbol{l}) \equiv \mathbf{u(r }+\, \boldsymbol{l}) - \mathbf{u(r)}$ and the magnetic field increment $\delta \mathbf{b(r,} \, \boldsymbol{l}) \equiv \mathbf{b(r +}\, \boldsymbol{l}
)
- \mathbf{b(r)}$. Note that these increments 
have 6-dimensional arguments, depending on both 
real (regular) space and lag space, but are three dimensional vectors.

Averages of functions of the increments in real space form the basis for the third order law. Note that from the definition of the turbulent fluctuations $\mathbf{u(r)}$ and $\mathbf{b(r)},$ $\langle \mathbf{u(r)} \rangle_r$ and $\langle \mathbf{b(r)} \rangle_r$ = 0. The second order structure functions are defined as $S_u(\boldsymbol{l}) \equiv \langle\, \mathbf{\lvert \delta u(r,}\,\boldsymbol{l}) \rvert^2\,\rangle_r$ and $S_b(\boldsymbol{l}) \equiv \langle\, \mathbf{\lvert \delta b(r,}\,\boldsymbol{l})\rvert^2\,\rangle_r$. The general second order structure function including both magnetic and velocity fluctuations is $S(\boldsymbol{l}) = S_u(\boldsymbol{l}) + S_b(\boldsymbol{l})$. In Alfv\'enic units, the structure functions have units of energy per unit mass. 

In this form, the structure functions are unwieldy because of their three
dimensional dependence on the vector lag $\boldsymbol{l}$. They are often averaged over 
direction (solid angle) to give a function of only the magnitude of lag
$l,$ as 
\begin{equation} \label{av-S}
\bar{S}(l) \equiv \langle S(\boldsymbol{l}) \rangle_{\Omega_l} = \frac{1}{4\pi}\;\int d\Omega_l\,S(\boldsymbol{l}),
\end{equation}
where $\Omega_l$ is the solid angle in $l$ space.

By expanding out the terms in the second order structure functions, the physical meaning becomes apparent. Examining $\bar{S}_u{(l)}$,
\begin{equation} \label{S-intuitive}
\bar{S}_u(\boldsymbol{l}) = 2\langle\;\langle\,\lvert \mathbf{u(r)}\rvert^2\,\rangle_r\;\rangle_{\Omega_l} - 2\langle\;\langle \mathbf{u(r} + \boldsymbol{l}) \cdot \mathbf{u(r)} \rangle_r\;\rangle_{\Omega_l},
\end{equation}
where we have used the fact that $\langle\,\lvert \mathbf{u(r}+\boldsymbol{l})\rvert^2\,\rangle_r = \langle\,\lvert \mathbf{u(r)}\rvert^2\,\rangle_r$ 
for homogeneous turbulence. 
The first term is four times the average energy per unit mass of velocity fluctuations in the system $4\,E_{uav}$ and the second term is twice the unnormalized autocorrelation function of $\mathbf{u(r)}$ which we denote as $\bar{R}_u(l)$. 
\begin{figure}
\begin{center}
\includegraphics[width=4in]{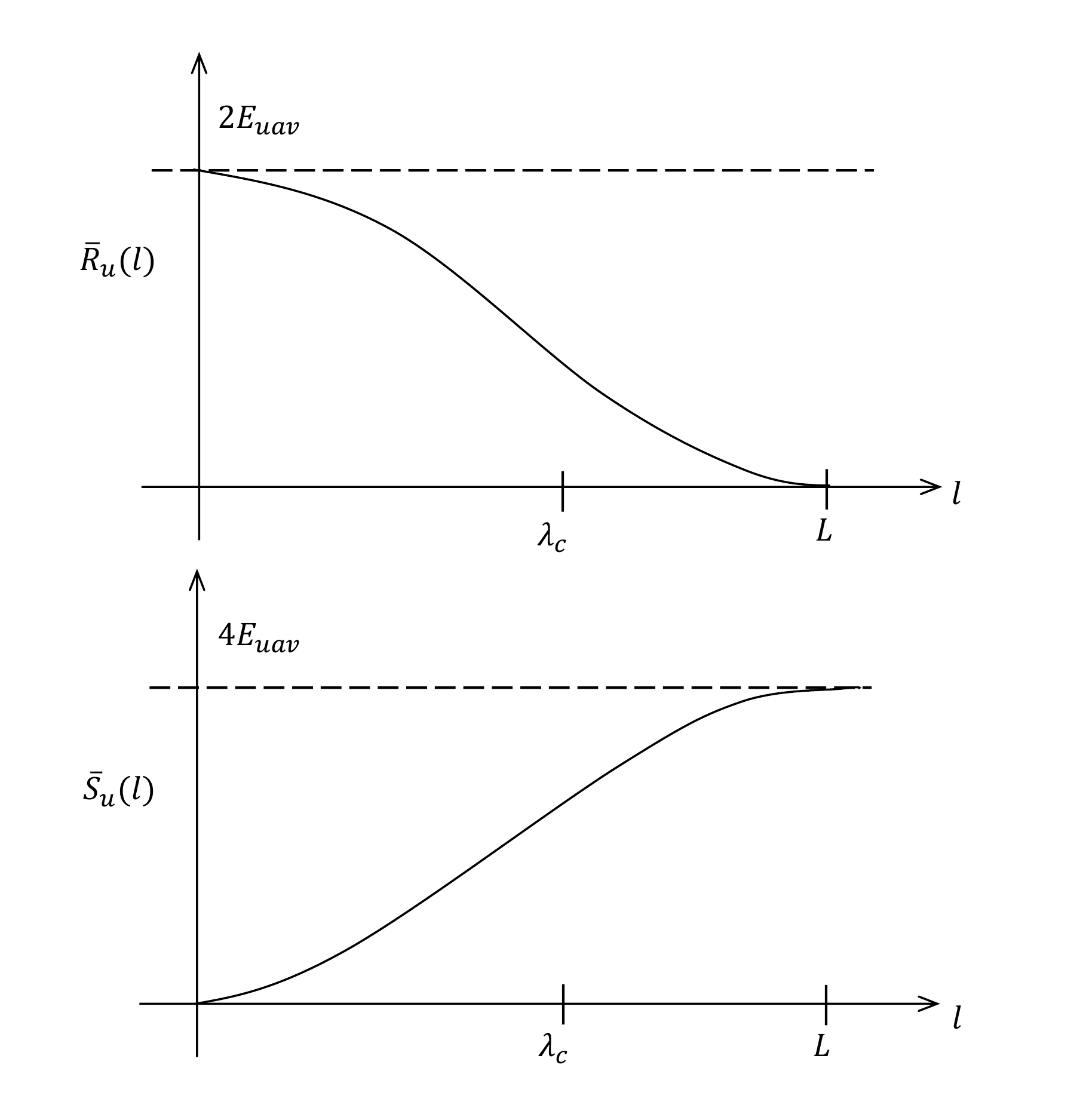}
\end{center}
\caption{\label{sample-R-S} Sample autocorrelation function $\bar{R}_u(l)$ and second order velocity structure function $\bar{S}_u(l)$. The correlation scale $\lambda_c$ is the integral scale $\int_0^\infty R_u(\ell) dl$ which is commonly estimated as the lag where $\bar{R}_u(l) = \bar{R}_u(0)/e$.} 
\end{figure}
Fig.~\ref{sample-R-S} shows representative functions of $\bar{R}_u(l)$ and $\bar{S}_u(l)$ for a turbulent system of size $L.$ $\bar{R}_u(l)$ peaks at the origin and then falls off over a length comparable to the correlation scale $\lambda_c$. $\bar{S}_u(l)$ is zero at the origin and gradually rises, reaching $4\,E_{uav}$ at the system size. The continuous rise of $\bar{S}_u(l)$ implies that it is a cumulative function of $l.$ An intuitive meaning of $\bar{S}_u(l)$ then presents itself as: \textit{Four times the average velocity field energy/mass in fluctuations of size between $0$ and $l$.} $\bar{S}_b(l)$ and $\bar{S}(l)$ have related physical meanings linked to the magnetic field energy and total energy, respectively.

Now that the physical meaning of the second order structure functions has been established, we address the question of how they vary in time. The behavior is governed by the von K\'arm\'an Howarth equation, which is derived by massaging the dynamical fluid equations at $\mathbf{r}$ and $\mathbf{r} + \boldsymbol{l}$ to create time derivatives of the structure functions~\citep{PolitanoPRE1998}. The resulting time derivatives are dependent on averages of higher order increments of velocity and magnetic fields as can be seen from the appearance of third-order structure functions in the von K\'arm\'an Howarth equation. 
In homogeneous hydrodynamic turbulence, and within a broad band inertial range, several terms in the von K\'arm\'an Howarth equation may be neglected and a third-order law emerges that 
gives an exact relationship between energy decay rate and the third-order structure function \cite{Kol41c}. This formalism was developed for 
MHD by Politano and Pouquet \cite{PolitanoPRE1998}. Recently, the 
von K\'arm\'an Howarth equation 
has been generalized to Hall-Magnetohydrodynamics in a non-isotropic form, namely, \cite{hellinger2018karman,ferrand2019exact}:


\begin{equation} \label{cascadeeqn_later}
 \frac{\partial S (\boldsymbol{l})}{\partial t} +\nabla_l \cdot \mathbf{Y}(\boldsymbol{l})+\frac{1}{2}\,\nabla_l \cdot \mathbf{H}(\boldsymbol{l})= 2D(\boldsymbol{l})  -4\epsilon,
\end{equation}
with $\mathbf{Y}(\boldsymbol{l})=\langle \delta \mathbf{u} \lvert \delta \mathbf{u} \rvert^2 +\delta \mathbf{u} \lvert \delta \mathbf{b} \rvert^2 -2\delta \mathbf{b}(\delta \mathbf{u} \cdot \delta \mathbf{b}) \rangle_r$~\cite{PolitanoPRE1998} and  $\mathbf{H}(\boldsymbol{l})=\langle 2\delta \mathbf{b}(\delta \mathbf{b} \cdot \delta \mathbf{j})-\delta \mathbf{j} \lvert \delta \mathbf{b} \rvert^2 \rangle_r$~\cite{hellinger2018karman}.
Both $\mathbf{Y}(\boldsymbol{l})$ and $\mathbf{H}(\boldsymbol{l})$ are mixed third-order structure functions that describe the cascade of energy in MHD and Hall MHD respectively. On the right hand side of equation~\ref{cascadeeqn_later}, $\epsilon$  is a constant independent of $\boldsymbol{l}$, while 
$D(\boldsymbol{l})$ is a lag dependent dissipation term. In the collisionless kinetic simulations in this study, the exact form of these terms in lag space are not known. However, for the scale filtered form, the 
analogous term is the filtered pressure-strain interaction \cite{yang2022pressure}.
In a system with kinematic viscosity $\nu$ and resistivity $\eta,$ these terms take the form $\epsilon=\nu \langle (\nabla \mathbf{u}: \nabla
\mathbf{u}) \rangle_r + \eta \langle \nabla \mathbf{b}: \nabla \mathbf{b} \rangle_r$ and $D(\boldsymbol{l}) = \nu
\nabla_l^2 S_u(\boldsymbol{l}) + \eta \nabla_l^2 S_b(\boldsymbol{l}). $ 

Before discussing the physical meaning of the terms of Eq.~\ref{cascadeeqn_later} we simplify them as we did with the structure functions by averaging over solid angle in lag space. As we did with Eq.~\ref{av-S}, we denote averages in solid angle by putting a ``$\bar{\;}$'' over variables. The von K\'arm\'an equation
becomes:
\begin{equation} \label{av-third-order}
\frac{1}{4}\,\frac{\partial \bar{S}(l)}{\partial t} + \frac{1}{4}\,\nabla_l \cdot \bar{\mathbf{Y}}_l(l) + \frac{1}{8}\, \nabla_l \cdot \bar{\mathbf{H}}_l(l) = \frac{1}{2}\,\bar{D}(l) - \epsilon,
\end{equation}
where $\bar{\mathbf{Y}}_l(l) = \hat{l}\,(\hat{l} \cdot \bar{\mathbf{Y}})$, $\bar{\mathbf{H}}_l(l) = \hat{l}\,(\hat{l} \cdot \bar{\mathbf{H}})$, and $\hat{l}$ is the unit vector along the radial direction in lag space. In writing these terms we have used that $\langle\,  \nabla_l \cdot \mathbf{Y(l)}\, \rangle_{\Omega_l} = \nabla_l \cdot \bar{\mathbf{Y}}(l)$ and 
$\langle\,  \nabla_l \cdot \mathbf{H(l)}\, \rangle_{\Omega_l} = \nabla_l \cdot \bar{\mathbf{H}}(l)$
~\cite{taylor2003recovering,wang2022strategies}.
Written explicitly $\nabla_l \cdot \bar{\mathbf{Y}}_l(l) = (1/l^2)\,\partial/\partial l\, (l^2\,\hat{l} \cdot \bar{\mathbf{Y}})$.
Note that the solid angle averaged vectors $\bar{\mathbf{Y}}(l)$ and $\bar{\mathbf{H}}(l)$ could in principle have nonzero components in lag space along the polar angle $\theta_l$ and the azimuthal angle $\phi_l$, but these components do not contribute to the direction averaged divergence. 
For more discussion of the 
role of direction averaging, see the references \cite{wang2022strategies}.

In studying the physical meaning of each term of Eq.~\ref{av-third-order}, it is tempting to draw analogies with the energy equation in electricity and magnetism, with $\partial \bar{S}(l)/\partial t$ analogous to the rate of change in time of the energy density of the fields, $\bar{\mathbf{Y}}_l(l)$ and $\bar{\mathbf{H}}_l(l)$ analagous to the Poynting flux, and the RHS analogous to the particle/fields energy exchange term $\mathbf{J \cdot E}$. In line with this analogy, Eq.~\ref{av-third-order} can be integrated over some volume of lag space with the divergence terms becoming surface integrals. However, $\bar{S}(l)$ does not represent a local energy density in lag space, but instead a volume integrated quantity from zero lag to $l.$ As such, integrating $\bar{S}(l)$ over a volume in lag space, although possible and sometimes mathematically expedient, does not have a clear physical interpretation. Related to this fact, as will be discussed shortly $\bar{\mathbf{Y}}_l(l)$ and $\bar{\mathbf{H}}_l(l)$ are not energy flux densities in lag space and using the term ``flux'' to describe them is potentially confusing. 

Note that in the following paragraphs discussing the physical meaning of Eq.~\ref{av-third-order}, for simplicity we so not separately treat the $\nabla_l
\cdot \bar{\mathbf{H}}_l(l)$ term. This term has
the same form as the MHD cascade term $\nabla_l \cdot \bar{\mathbf{Y}}_l(l)$
and the inertial range can be thought of as including both an MHD cascade
range and a Hall cascade range \cite{BandyopadhyayEA20-Hall}.

\begin{figure}
\begin{center}
\includegraphics[scale=0.5]{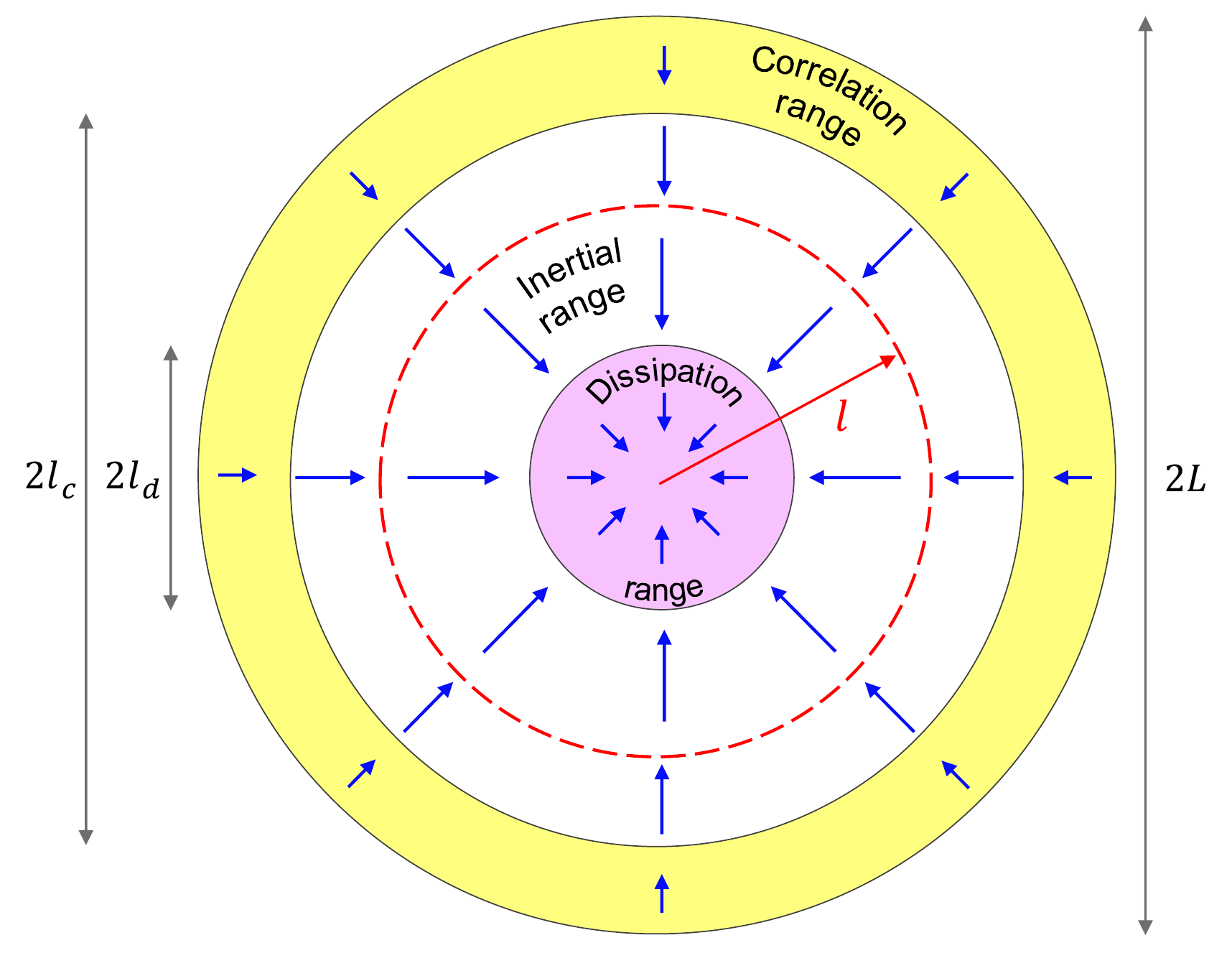}
\end{center}
\caption{\label{schematic-lagspace} Schematic showing idealized lag space
of decaying turbulence in an isotropic case. Dissipation range $(l < l_d)$
in pink, inertial range ($l_d < l < l_c$) in white, and correlation range
($l_c < l < L$)  in yellow. Energy/mass flux density leading to cascade shown
with blue arrows. In red, a representative lag sphere of radius $l$ is shown. 
} 
\end{figure}

To aid in the discussion of the physical meaning of the terms of the third order law, a much simplified schematic of a decaying turbulent system in lag space is shown in Fig.~\ref{schematic-lagspace}. For simplicity the system depicted is isotropic. The system spans lags from $l = 0$ to the size $l = L.$ Three physical ranges in lag space are denoted by colors, the correlation range (also known as the energy range or the energy containing range) 
in yellow, the inertial range in white, and the dissipation range in pink, with the (approximate) boundaries at the dissipation lag $l_d$ and the correlation lag $l_c.$ The transfer or cascade of fluctuation energy to smaller lag is shown with blue arrows. A representative ``lag sphere'' of radius $l$ is shown in red inside the inertial range. 

We choose to examine decaying turbulence in an isolated system in lag space. That is, no energy transfer occurs through the boundary at $l = L.$ In addition, we assume that the Reynold's number is large enough in the system such that the total energy/mass in the system $\bar{S}(L)$ is changing very slowly relative to the 
timescales for the inertial range cascade. This means that the time variation of $\bar{S}(l)$ in the inertial and dissipation ranges is very small. 

The von K\'arm\'an Howarth equation
(Eq.~\ref{av-third-order}) physically represents the flow of energy in the lag space shown in Fig.~\ref{schematic-lagspace}. 
The term  $(1/4)\partial\bar{S}/\partial t$ is the rate of change of energy/mass inside a lag sphere of radius $l.$ $(1/4)\,\nabla_l \cdot \bar{\mathbf{Y}}_l(l)$ is the MHD energy/mass transfer through the lag sphere $l$ (with $(1/8)\,\nabla_l \cdot \bar{\mathbf{H}}_l(l)$ being due to Hall physics).

The right hand side of Eq.~\ref{av-third-order} is the average dissipation rate of energy/mass inside a lag sphere of radius $l.$ For analysis purposes focussed on the rate of energy cascade in the inertial range, it has been broken up into two terms. $\epsilon$ is the average dissipation rate of energy/mass in the system, as is clear from its definition in the case of fluid $\nu$ and $\eta$. It is often called the ``cascade rate'' in the literature, because in an idealized inertial range the energy cascade rate is equal to $\epsilon$ as will be discussed concerning Eq.~\ref{eq:inertial}.  The other term $\bar{D}(l)/2$ on the RHS of Eq.~\ref{av-third-order} is clearly related to dissipation but is usually not discussed in the literature, primarily because it is very small in the inertial range. $\bar{D}(l)/2$ is strange at first glance, because it is \textit{positive}, meaning that a dissipative term at face value acts to increase the energy in time. However, as we shall see it represents the rate of energy/mass dissipated outside of the lag sphere of radius $l$ and acts as a counterweight to the total dissipation $\epsilon.$

\begin{figure}
\begin{center}
\includegraphics[scale=0.7]{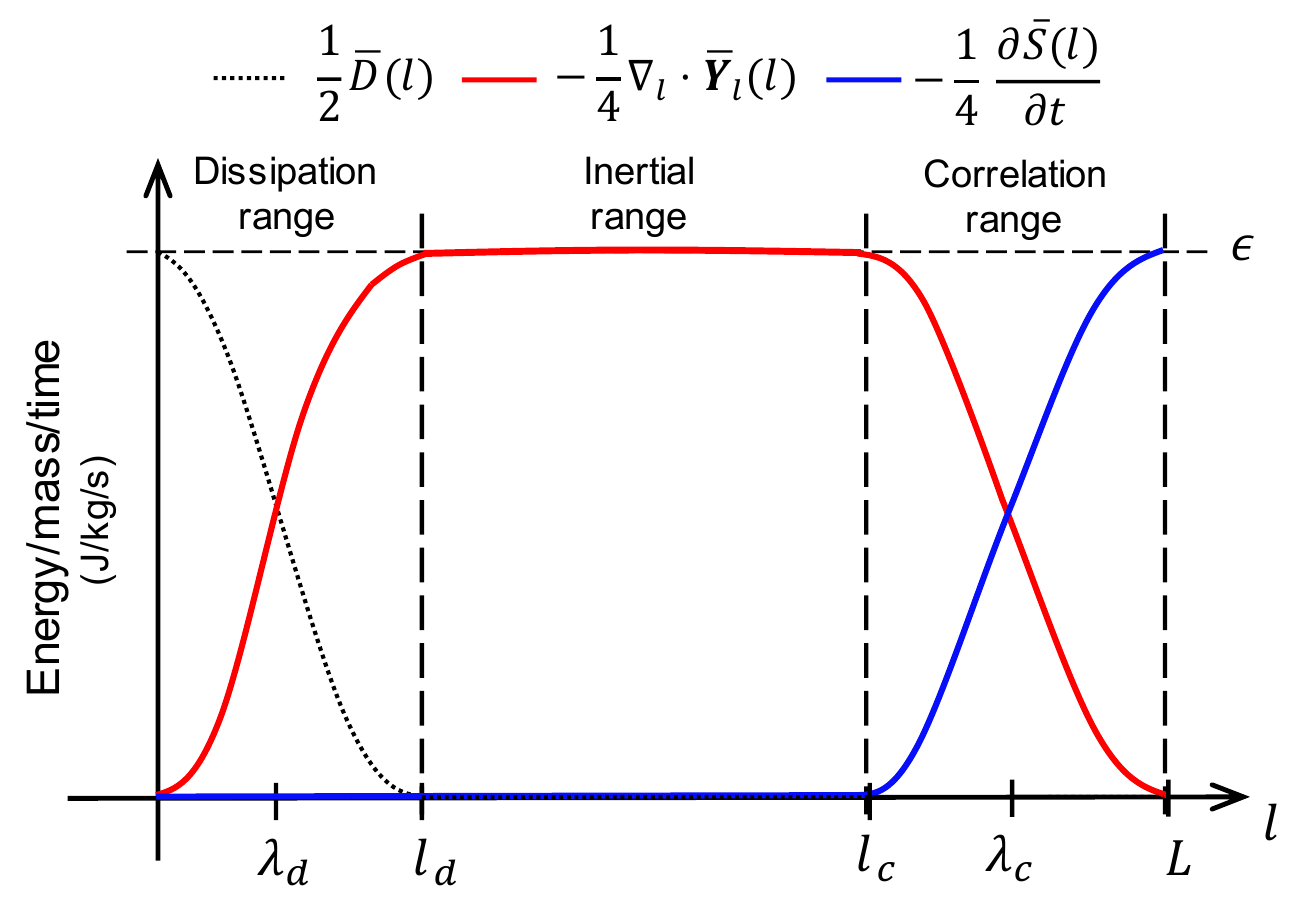}
\end{center}
\caption{\label{third-order-terms}Idealized schematic plot of different terms of third order law versus $l$, where the Hall cascade term  $(1/8)\,\nabla_l \cdot \bar{\mathbf{H}}_l(l)$ has been ignored to simplify the discussion. Horizontal dashed line is $\epsilon.$ Representative values of the correlation scale $\lambda_c$ and the Kolmogorov scale $\lambda_d$ are shown. 
} 
\end{figure}

We now examine the properties of Eq.~\ref{av-third-order} in the different regions of lag space shown in Fig.~\ref{schematic-lagspace}, highlighting the associated physics. To assist this analysis in Fig.~\ref{third-order-terms} is shown an idealization of the variation of the 
terms in  Eq.~\ref{av-third-order} 
in lag space (we temporarily ignore the Hall cascade term). 
Representative values of the correlation scale $\lambda_c$ and Kolmogorov scale $\lambda_d$ are shown for reference. The value of the total dissipation rate of energy/mass $\epsilon$ is drawn as the horizontal dotted line. At the outer boundary $l = L,$ $(1/4)\,\partial \bar{S}/\partial t = -\epsilon,$ meaning that the rate of change of energy/mass in the entire system must equal the total dissipation rate, which is required if the system is isolated.  

In the inertial range,  the influences of 
$\partial\bar{S}/\partial t$ and $\bar{D}$ terms are small, and we have 
an inertial range in which $(1/4)\,\nabla_l \cdot \bar{\mathbf{Y}}_l(l) = -\varepsilon.$ 
Physically, this is simply the statement that 
when $\bar{S}(l)$ is a constant 
in time in the inertial range, the energy transfer into a lag sphere of radius
$l$ must balance the total dissipation of energy within the sphere. 
In the literature, the third order law in the inertial range is often integrated over the \textit{volume} of the lag space sphere, giving $\frac{1}{4}\,  (4\pi l^2)\,(\hat{l} \cdot \bar{\mathbf{Y}}_l(l)) = -\frac{4}{3}\pi l^3 \epsilon$, 
thus arriving $\bar{\mathbf{Y}}_l(l) = -\frac{4}{3}\,\epsilon\,l\,\hat{l}$, 
equivalent to the familiar scalar relation
$\bar{Y}_l(l) = -\frac43 \epsilon l$. 

Unlike the meaning of Poynting flux, the surface integral $\int \bar{\mathbf{Y}}_l \cdot \mathbf{dA}_l$ is not the rate of energy/mass transfer through a surface in lag space. Similarly, $\bar{\mathbf{Y}}_l(l)$ is often called the ``MHD turbulent cascade flux'' in the literature but physically \textit{does not represent an energy/mass flux density}. If it were truly a flux density in lag space, energy/mass conservation in the inertial range would require $\nabla_l \cdot \bar{\mathbf{Y}}_l(l) = 0$. For that reason we choose not to use the term ``flux'' to describe $\bar{\mathbf{Y}}_l(l)$ (nor $\bar{\mathbf{H}}_l(l)$), but instead simply refer to their divergence as the transfer rate or cascade rate of energy/mass in lag space. To determine the $\bar{\mathbf{Y}}_l(l)$ in the inertial range, it is more physically meaningful to directly solve the third order law differential equation in the inertial range,
\begin{equation} \label{eq:inertial}
\frac{1}{4 l^2}\; \frac{\partial}{\partial l}\,\left[l^2 (\hat{l} \cdot \bar{\mathbf{Y}}_l(l)  )  \right] = - \epsilon,
\end{equation}
which gives the same solution mentioned above. We stress that the actual solid angle averaged energy/mass flux density along $\hat{l}$ in the inertial range is $-\hat{l}\,\epsilon/(4\pi\,l^2),$ which as expected has a zero divergence in the inertial range. 

In the correlation range in Fig.~\ref{third-order-terms}, S is decreasing in time to feed the cascade and ultimate dissipation of energy/mass. $(1/4)\,\partial\bar{S}/\partial t$ is equal to $\epsilon$ at $l = L$ and becomes steadily smaller with decreasing $l,$ reaching 0 at the inner boundary of the correlation range. Conversely, the energy/mass transfer $\nabla_l \cdot \bar{\mathbf{Y}}_l(l)$ is 0 at $l = L$ and increases with decreasing $l.$ In the middle of this range, the time change term is of the same order as the transfer term. Examining a lag sphere in the middle of the correlation range, the energy drained from $\bar{S}(l)$ at larger lags must be transported towards the inertial range raising the value of the cascade term. Conversely, the change of energy in the lag sphere has decreased because it now contains less of the correlation range where $\partial \bar{S}/\partial t$ is nonzero.  

The dissipation range ($l < l_d$) is characterized by non-negligible $\bar{D}(l)$ and $\partial\bar{S}/\partial t = 0$ with Eq.~\ref{av-third-order} simplifying to (ignoring the Hall transfer term)
\begin{equation} \label{diss-range}
 \frac{1}{4}\,\nabla_l
\cdot \bar{\mathbf{Y}}_l(l) = \frac{1}{2}\,\bar{D}(l) - \epsilon.
\end{equation}
As seen in Fig.~\ref{third-order-terms}, as $l$ decreases from $l_d,$ the cascade term drops while $\bar{D}(l)$ rises. Drawing a lag sphere in the central region where $\nabla \cdot \bar{\mathbf{Y}}_l(l) \sim \bar{D}(l),$ the physical meaning is clear. To maintain the time constancy of $\bar{S}(l)$, any dissipation occurring inside the lag sphere (RHS of Eq.~\ref{diss-range}) must be balanced by the transfer of energy into the sphere. Since $\epsilon$ is the total dissipation occuring in the system, $\bar{D}(l)/2$ must be the energy dissipation occurring \textit{outside} the lag sphere. At $l =0,$ there is no energy transfer in the lag sphere and thus $\bar{D}(0) = \epsilon,$ meaning that all dissipation is occurring outside of a 0 radius lag sphere.


In summary, the key points we wish to stress in this review section are:  
\begin{itemize}
\item The lag $l$ represents a length of variation of the fluctuating velocity $\mathbf{u}$ and magnetic field $\mathbf{b}$. Discussions of the properties of turbulence in lag space are directly related to more well known spectral analysis through the relation $\boldsymbol{l} \sim \mathbf{k}/k^2$.
\item The direction-averaged second order structure function $\bar{S}(l)/4$ is physically the fluctuation energy per unit mass contained in lag space spanning lags of 0 up to lags of $l.$ 
\item $\epsilon$ is the total rate of dissipation of energy per unit mass dissipated in the system. In the inertial range it is equal to the cascade rate of energy per unit mass. 
\item Even though $\bar{\mathbf{Y}}(l)$ and $\bar{\mathbf{H}}(l)$ are often called the ``turbulent cascade flux'', they \textit{do not physically represent} an energy flux density in lag space. In the inertial range of a quasisteady cascade where $\partial\bar{S}/\partial t = 0$, the divergence in lag space of an energy flux density must be zero. We now move on to describe the simulations and results.
\end{itemize}

\section{\label{sec:simulations}Simulations}

For this study, five kinetic 2.5D particle in cell (PIC) simulations are performed with the double Harris sheet equilibrium and various guide fields. Length is normalized to the ion inertial length $(d_i = \sqrt{m_i c^2/4\pi n_0 e^2})$, time is normalized to the inverse of ion cyclotron frequency [$\omega_{ci}^{-1} = (eB_0/m_i c)^{-1}$], and speed is normalized to the ion Alfv\'en speed ($v_A$). The magnetic field is normalized to $B_0$, number density is normalized to $n_0$, electric field is normalized to $E_0=v_A B_0/c$ and temperature is normalized to $T_0=\frac{1}{2}m_iv_A^2$.

All the simulations are of size $L_x=L_y=204.8d_i$ with grid spacing of $\delta x =0.05$ and total grid points of $4096^2$. The guide field is varied as $B_g=\left[ 0,0.1,0.5,1, 2\right]B_0$. The speed of light $c=15v_A$ and the half width of the current sheets is $3d_i$. More details of the simulation are listed in Table \ref{tab:table:sim1} (See Adhikari et al. \cite{adhikari2020reconnection,adhikari2021magnetic} for R1).  As a cross check of resolution, an additional $R5$ simulation was performed with a higher resolution $( \delta x =0.025) $; no significant change in the results were found so this simulation is not included in this analysis. 

\begin{table}[h]
\fontsize{9}{12}\selectfont
\caption{\label{tab:table:sim1}Simulation details: background density $n_b$, mass (m) of ions(i)/electrons(e), temperature ($T$), out of plane guide field $B_g$, reconnecting magnetic field $B_r$, and particles per grid (ppg).}
\begin{ruledtabular}
\begin{tabular}{ccccccccc}
Run & $L_{x}=L_{y}$ & $n_b$ & $m_e/m_i$ & $T_e/T_i$  & $B_g$ & $B_r$ & $\beta_e/\beta_i$ &$ppg$ \\
\hline\hline
R1 & $204.8d_i$ & $0.2$ & $0.04$ & $0.25/1.25$ & $0$ & $1$ & $0.1/0.5$  &$100$ \\
\hline
R2 & $204.8d_i$ & $0.2$ & $0.04$ & $0.25/1.25$  & $0.1$ & $1$ & $\sim 0.1/0.5$ &$100$ \\
\hline
R3 & $204.8d_i$ & $0.2$ & $0.04$ & $0.25/1.25$  & $0.5$ & $1$ & $0.09/0.45$&$100$ \\
\hline
R4 & $204.8d_i$ & $0.2$ & $0.04$ & $0.25/1.25$  & $1$ & $1$ & $0.07/0.35$ &$100$ \\
\hline
R5 & $204.8d_i$ & $0.2$ & $0.04$ & $0.25/1.25$  & $2$ & $1$ & $0.044/0.22$ &$100$ \\
\end{tabular}
\end{ruledtabular}
\end{table}

\section{\label{sec:results}Results}
\subsection{Overview of the simulations}
\begin{figure}
\includegraphics[scale=0.7]{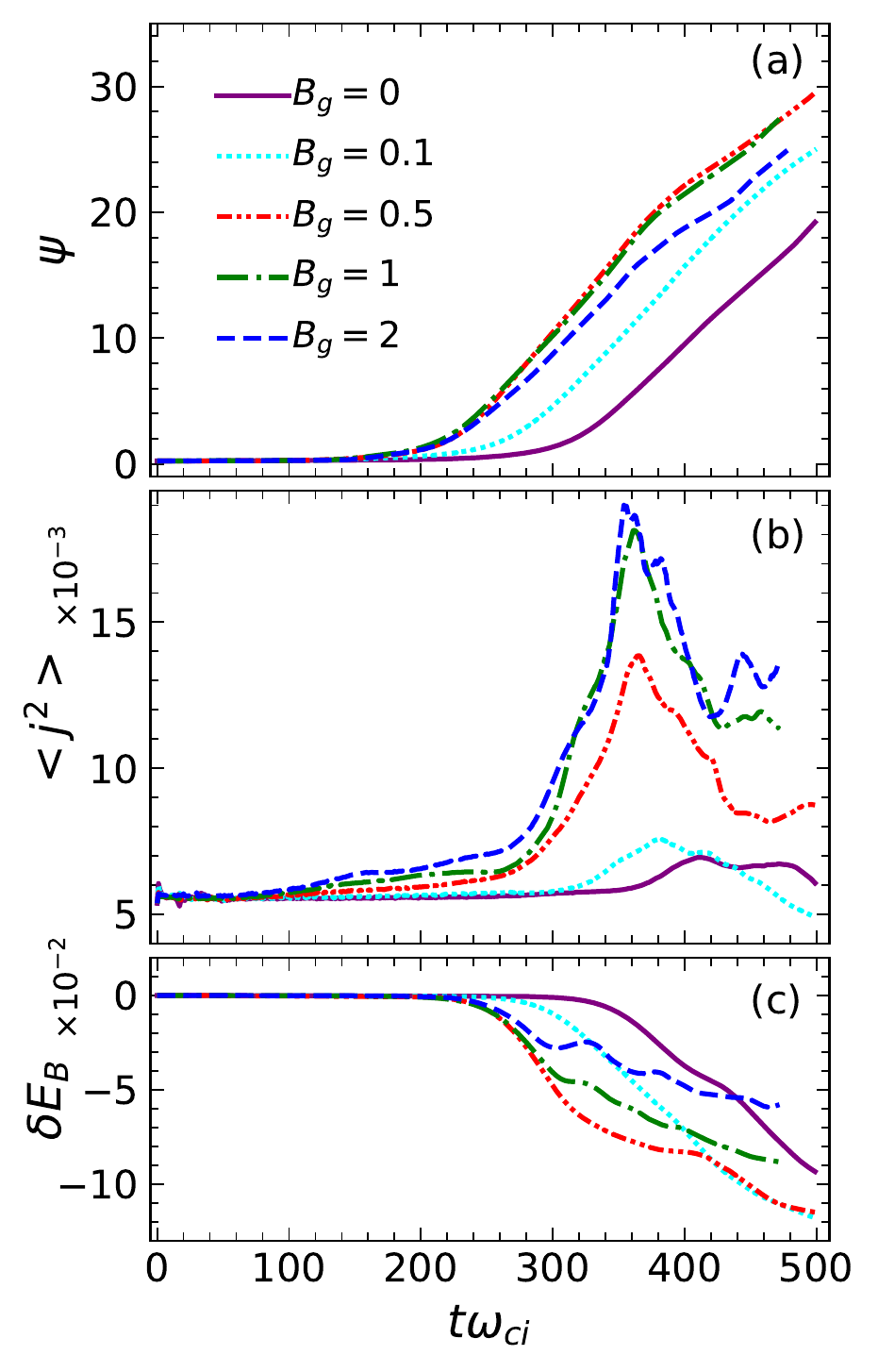}
\caption{\label{fig:psi_jsq_recon} 
Time evolution of the (a) reconnected flux $\psi$, (b) mean square current $<j^2>$, and (c) change in magnetic energy $E_B = <\frac{B^2}{2} > $ for all simulations. }
\end{figure}

\begin{figure}
\includegraphics[scale=0.7]{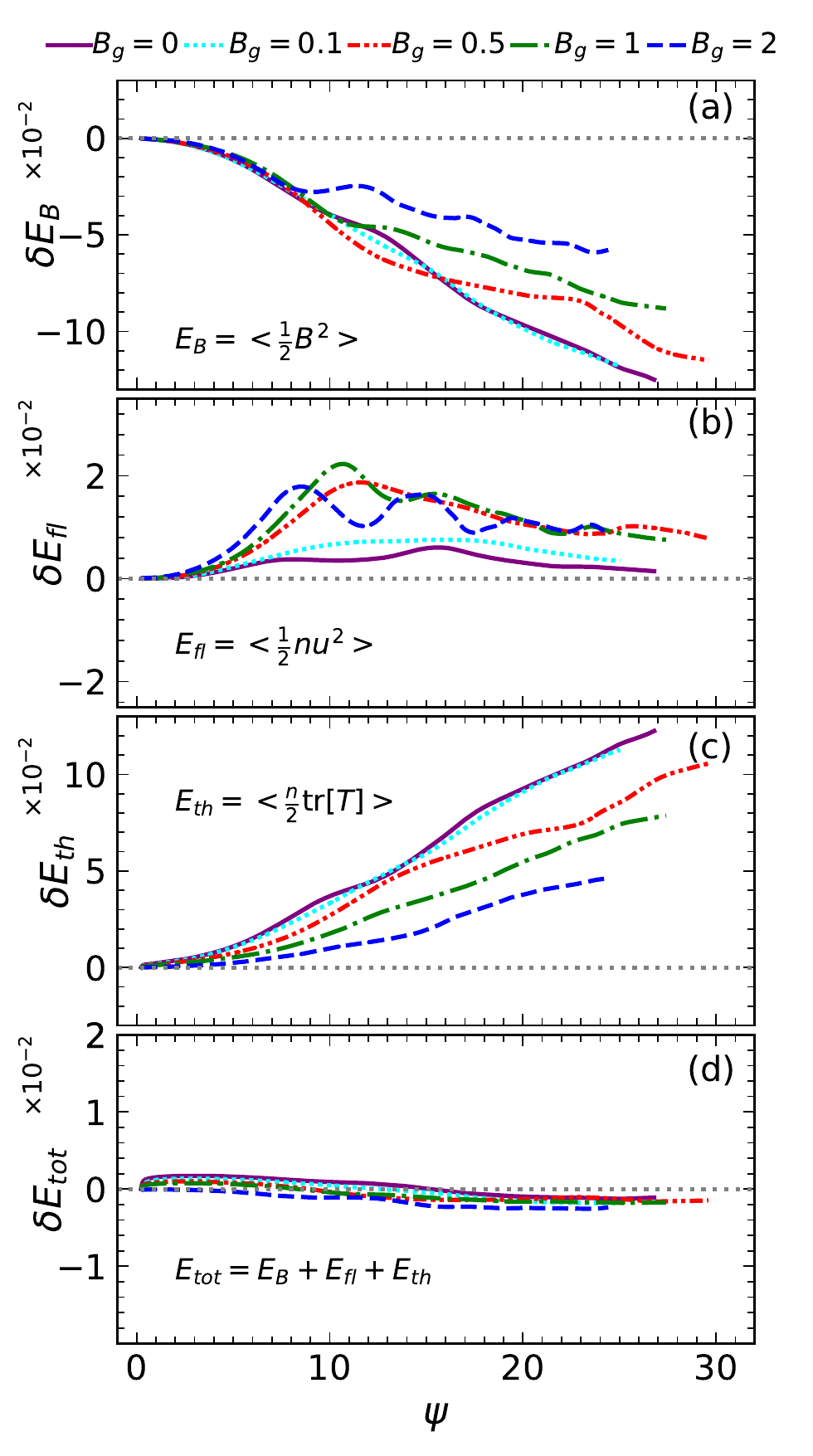}
\caption{\label{fig:energyrecon} 
Change in different forms of energies versus the reconnected flux: $E_B$ is the magnetic energy, $E_{fl}$ is the total bulk flow energy (ion plus electron), $E_{th}$ is the total thermal energy (ion plus electron), and $E_{tot}$ is the total energy.}
\end{figure}

We start the analysis with an overview of the simulations. Figure~\ref{fig:psi_jsq_recon}a shows the time evolution of the reconnected magnetic flux $\psi$~\cite{loureiro2009turbulent} and the mean square current $\langle j^2\rangle$. The average reconnection rate $\partial \psi/\partial t$ of both the top and bottom current sheet is similar for all the simulations, but the onset of reconnection differs substantially. The reconnection onset time for antiparallel reconnection $B_g = 0$ is around 100 $t\omega_{ci}$ later than the case with earliest onset time. The onset time decreases with increasing $B_g$, but appears to reach a minimum value 
at guide fields lower than $B_g = 0.5$. Presumably the larger $B_g$ cases better confine the electrons leading to a narrower current sheet, which allows reconnection to onset more quickly. Once reconnection initiates, however, the $B_g = 2$ case does exhibit a slightly slower reconnection rate than the fastest cases, which is consistent with other PIC simulations.~\cite{swisdak2005transition}

Consistent with the idea that a stronger guide field better confines the electrons, in Fig.~\ref{fig:psi_jsq_recon}b the peak value of the mean square current steadily increases with guide field, saturating around $B_g = 1.0$. The faster onset of reconnection with the larger $B_g$ cases is clearly evident in the time delay between different peaks of $\langle j^2\rangle$. This time delay complicates the analysis of the change of energies in the system. As an example, in Fig.~\ref{fig:psi_jsq_recon}c, the change in average magnetic energy $E_B = \langle B^2/2 \rangle$ behaves very differently in time for the different guide fields. 

To cross-compare more effectively, in Fig.~\ref{fig:energyrecon} are shown the change in different energies in the system (mean values) versus the reconnected flux. Note that all simulations show excellent energy conservation in panel~(d), with the total change in energy much smaller than any of the constituent parts. Until $\psi \approx 10$, the change in magnetic energy is nearly identical between all of the simulations. Interestingly, however, the change in bulk flow energy (b) diverges much earlier, and by $\psi = 10$ the larger guide field cases have substantially more flow energy; this larger flow energy is likely due to the higher reconnection outflow velocity in guide field versus antiparallel reconnection~\cite{haggerty2018reduction}. On the other hand, there is significantly more heating for small $B_g$ in (c). Clearly, the presence of the guide field is strongly affecting the fate of the magnetic energy released, with the larger guide field cases showing significantly less heating but more bulk flow energy~\cite{phan2014ion}.

For the reconnected flux $\psi > 10$ in Fig.~\ref{fig:energyrecon}, the simulations diverge from each other, even for the change in magnetic energy in panel (a); while the lowest guide field cases continue to show a steady decline, the reduction of magnetic energy in time becomes slower for larger guide fields. For the largest guide field cases, since  much of the magnetic energy is exchanged into flow energy, the change in flow energy in panel (b) decreases. Interestingly, for the strong guide field cases, $\delta E_{th}$ continues to increase, implying that this flow energy is being converted to thermal energy. One possible explanation is that at these later times, the exhaust bulk flow is compressing in the large magnetic island, converting to thermal energy. The very low guide field cases on the other hand, continue to show robust heating throughout the simulation.  

In analyzing the turbulence-like properties of these simulations, there is ambiguity associated with which time of the simulation is the most appropriate. While the time of peak $\langle j^2\rangle$ is one possibility, the clear organization of energy in Fig.~\ref{fig:energyrecon} associated with the change in reconnected flux is revealing. We study each guide field case when the reconnection flux $\psi \approx 15,$ which is near the time of peak $\langle j^2\rangle$ for all cases. Note that we have also performed these analyses at the time of peak $\langle j^2\rangle$, and found similar results. 

\begin{figure*}
\includegraphics[scale=0.6]{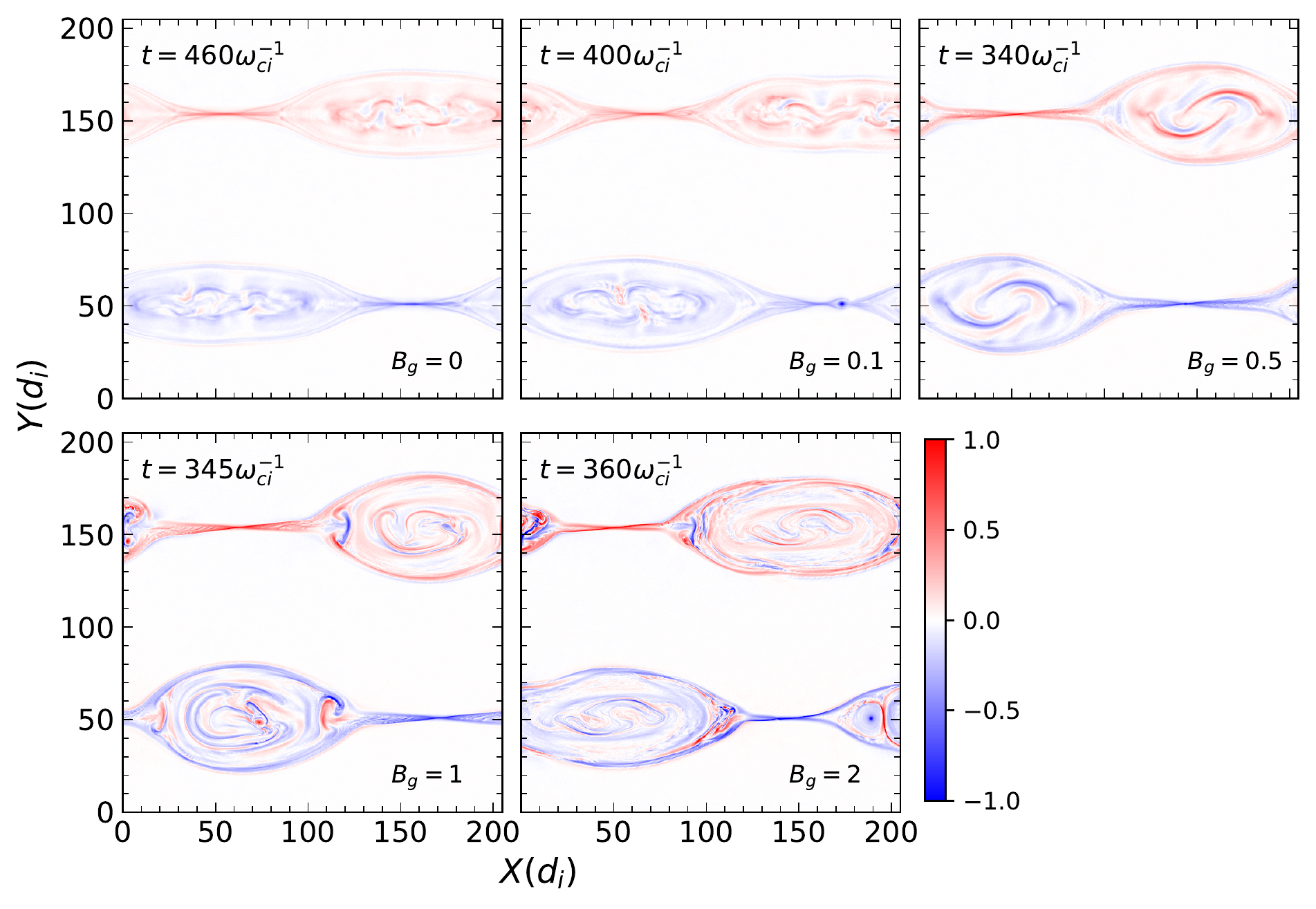}
\caption{\label{fig:jzreconn}Out of plane current $j_z$ for each guide field case when the amount of reconnected flux $\psi \approx 15$.}
\end{figure*}

Fig. \ref{fig:jzreconn} shows the out-of-plane current ($j_z$) density for all of the five runs when $\psi \approx 15$. The corresponding time (in $\omega_{ci}^{-1}$) for each run is included on each panel. For low guide fields, the current sheet in the vicinity of the x-line shows the extended but symmetric electron diffusion region seen previously~\cite{karimabadi2007multi,shay2007two}. For the stronger guide field cases, the current sheet is shifted towards the separatrices~\cite{swisdak2005transition,goldman2011jet,egedal2013review}. The currents near the x-line and in the magnetic island are clearly more intense and narrower for larger guide fields. The shape of the magnetic islands clearly show the role that compressibility is playing. For very low guide field, the magnetic islands are longer and narrower because the plasma compresses easily inside of them. For strong guide field cases, the islands are more round because the magnetic pressure at least partially inhibits compression of the plasma. For $B_g = 1$ and $2$, there are very intense currents at the end of the magnetic islands where flow is slowing down.
This effect may be due to the  ``backpressure'' as the plasma exhaust jets run into the growing magnetic islands; see Ref. \cite{matthaeus1986turbulent}.

\subsection{Energy spectrum: Electromagnetic field and Ohm's law}
\begin{figure*}
\includegraphics[scale=0.6]{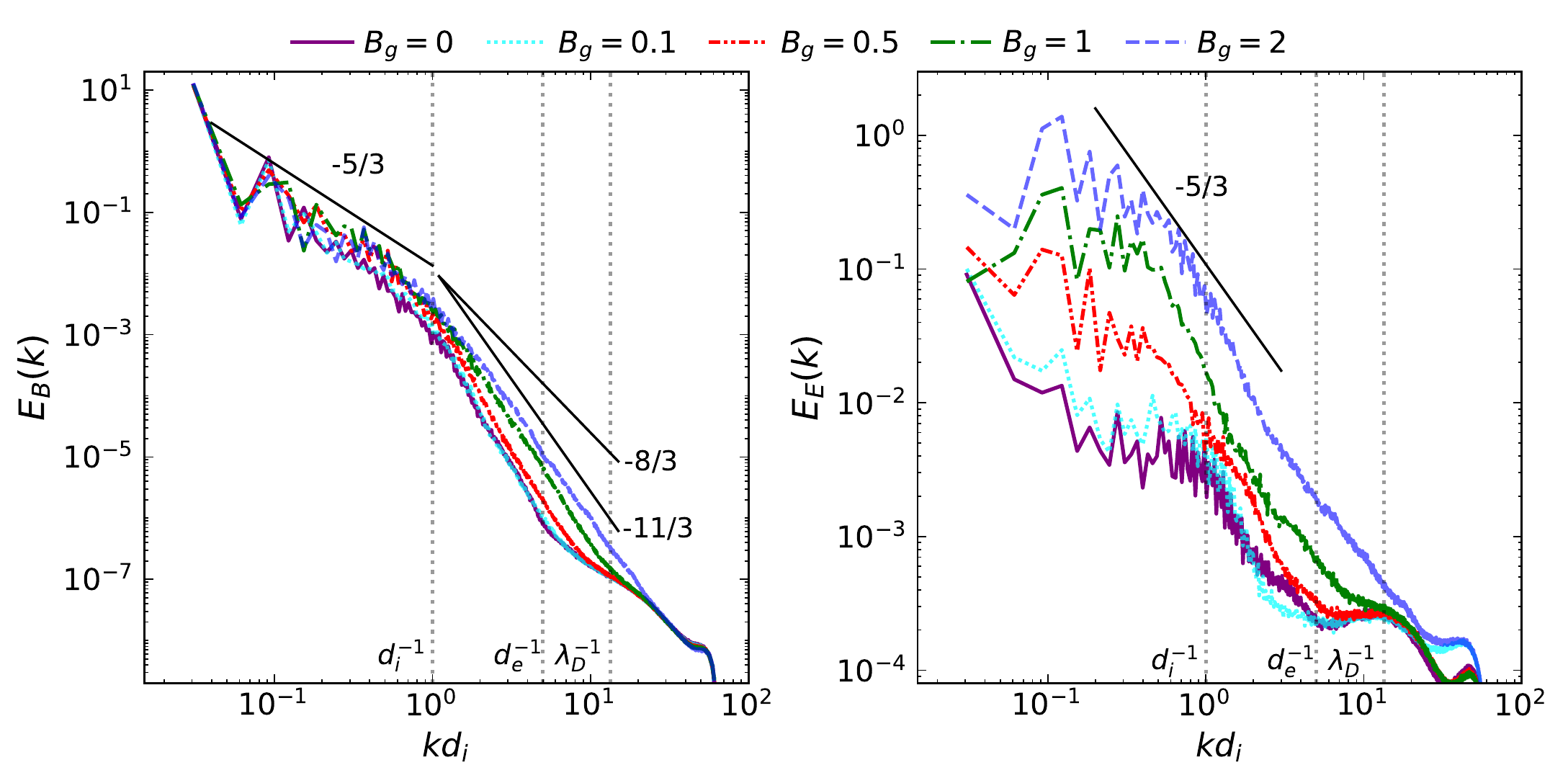}
\caption{\label{fig:magnetic_electric_spectra} Energy spectrum of the magnetic field (left) and electric field (right) at the time of analysis when the reconnected flux is about equal in all the simulations. Solid lines of slopes $-5/3$, $-8/3$, and $-11/3$ are drawn for reference. The vertical dotted lines represent the wavenumbers corresponding to the ion inertial length ($d_i$), electron inertial length ($d_e$) and the debye length ($\lambda_D$).}%
\end{figure*}

Having established the basic properties of the kinetic PIC simulations with varying guide field, it is natural to study whether a guide field changes the spectral and energy-transfer properties of reconnection relative to the antiparallel case ~\cite{adhikari2020reconnection,adhikari2021magnetic}.
Fig.~\ref{fig:magnetic_electric_spectra} shows the omni-directional magnetic (left) and electric (right) field spectra for all guide field cases. The magnetic spectra show a slope roughly consistent with $-5/3$ in the inertial range for all guide fields. 
Steeping from this value begins as $k$ increases beyond about $kd_i \sim 0.4$.
In the kinetic range, $(d_i^{-1} < k < d_e^{-1})$, the lower guide field cases have steeper spectral slopes. In this region the $B_g = 2$ case has a slope roughly consistent with -11/3.~\cite{smith2006dependence,sahraoui2009evidence,leamon1998observational}

The spectrum of the electric field (Fig.~\ref{fig:magnetic_electric_spectra}) exhibits more variation with changing guide field than the magnetic spectra. For $B_g = 0$ the electric spectrum has an $\approx 0$ slope for $0.1 \lesssim k\,d_i \lesssim 1.$ Conversely in this same region the $B_g = 2$ case has a negative slope with magnitude somewhat less than $5/3$. For very strong guide fields, MHD turbulent flows perpendicular to the guide field field $\mathbf{B}_0 = B_0\,\hat{z}$ are given by $\mathbf{u}_\perp \approx c\,\mathbf{E_\perp \times B}_0/B_0^2,$ where ``$\perp$'' denotes perpendicular to $\mathbf{B}_0;$ since the omnidirectional spectrum of $u_\perp$ has a power-law slope of $-5/3,$ by necessity $E_\perp$ will also have a slope of $-5/3$~\cite{bale2005measurement,chen2012density,stawarz2021comparative,matteini2017electric}. It is clear then, that as the guide field is increased, the electric spectrum for the range $0.1 \lesssim k\,d_i \lesssim 1$ is approaching but not quite reaching a slope of $-5/3.$ Consistent with this idea also, the electric field spectrum for $B_g = 2$ is dominated by the $E_x$ and $E_y$ components (not shown). 

The $B_g = 0$ case has a very different behavior. The electric field 
spectra shows a steep drop at the largest scales ($k\,d_i \lesssim 0.1$), but then flattens out in the inertial range $(0.1 \lesssim k\,d_i \lesssim 1).$ A clue to the reason for this flattening is that in this region $E_y$ plays the dominant role. Antiparallel reconnection is known to generate a large magnitude normal electric field approximately along $\hat{y}$, which extends for large distances along the separatrices and can have thicknesses much larger than $d_i$ ~\cite{shay1998structure,arzner2001magnetotail}. This global $E_y$ structure both dominates the omnidirectional spectra and creates a near $0$ slope.
For $k\,d_i > 1,$ in all cases there is a steepening of the electric field power spectra (Fig.~\ref{fig:magnetic_electric_spectra}).\cite{matteini2017electric,gonzalez2019turbulent} The spectra flatten out at very large $k,$ likely due to intrinsic noise in the PIC method associated with the finite particles per grid. 

\begin{figure*}
\includegraphics[scale=0.55]{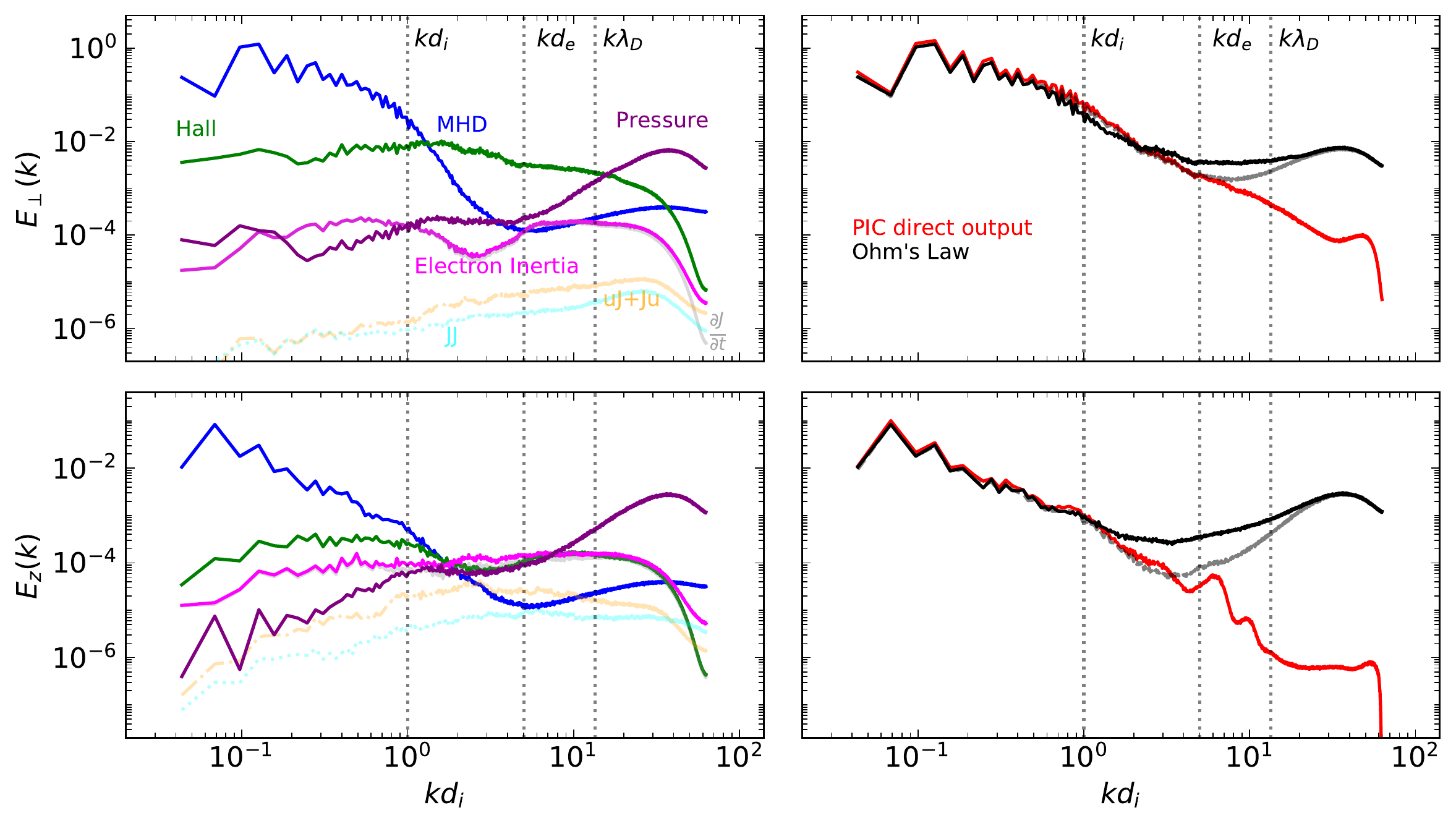}
\caption{\label{fig:ohmsspectra_g2_new}
Top panel (Left): Energy spectrum of different components of perpendicular electric field ($E_\perp$) as seen in the Ohm's law for Run R5 (\textbf{$B_g=2$)}. (Right) The sum of all the terms in the left (black), and energy spectrum of perpendicular electric field directly from the simulation(red). Bottom panel (Left): Energy spectrum of the different components of parallel electric field ($E_z$). (Right) The sum of all the terms in the left (black), and energy spectrum of parallel electric field (red). The gray curve drawn on the right panel is the spectrum of the sum of the individual terms, while the vertical dotted lines represent wavenumbers corresponding to the ion inertial length ($d_i$), electron inertial length ($d_e$) and the debye length ($\lambda_D$).}%
\end{figure*}

The electric field contains a wealth of information regarding the different physics acting at different length scales. In order to directly study the interplay of different physical scales, we explore the spectra of the different terms in the generalized Ohm's law. We begin with a focus on the $B_g = 2$ case and later compare all of the different guide field
cases.

The generalized form of the Ohm's law appropriate for collisionless plasma can be written as:
\begin{equation}
    \mathbf{E} = -\mathbf{u} \times \mathbf{B} +\frac{1}{n} \mathbf{J} \times \mathbf{B} -\frac{1}{n}\nabla \cdot \textbf{P}_e +\frac{d_e^2}{n}\left[\frac{\partial \mathbf{J}}{\partial t}+\nabla \cdot \left(\mathbf{u}\mathbf{J}+\mathbf{J}\mathbf{u}-\frac{\mathbf{J}\mathbf{J}}{ne}\right)\right],
\label{eqn:ohmslaw}
\end{equation}
where $\mathbf{u}=(1-\mu)\mathbf{u_i}+\mu \mathbf{u_e}$ is the single fluid bulk velocity, $\mu=m_e/(m_i+m_e)$, and ${\bf u}_i$, ${\bf u}_e$ are the mean velocities of ions and electrons respectively. $\mathbf{E}$, $\mathbf{B}$ are the electromagnetic fields, $\mathbf{J}$ is the electric current density, while $\textbf{P}_e$ is the electron pressure tensor. The pressure tensor can be further written as  $\textbf{P}_{e}=p_e + \Pi_e$, where $p=\frac{1}{3}\; \sum_i P_{ii}$ and $\Pi_{ij}=P_{ij}-p\delta_{ij}$ are the isotropic and anisotropic decomposition of the pressure tensor respectively. The first term on the right hand side 
of Eq. \ref{eqn:ohmslaw} 
is the MHD term (induction term), the second term is the Hall term (significant at Hall scales), the third term is the pressure term, and the remaining terms are associated with electron inertia.


\begin{figure*}[ht]
\includegraphics[scale=0.5]{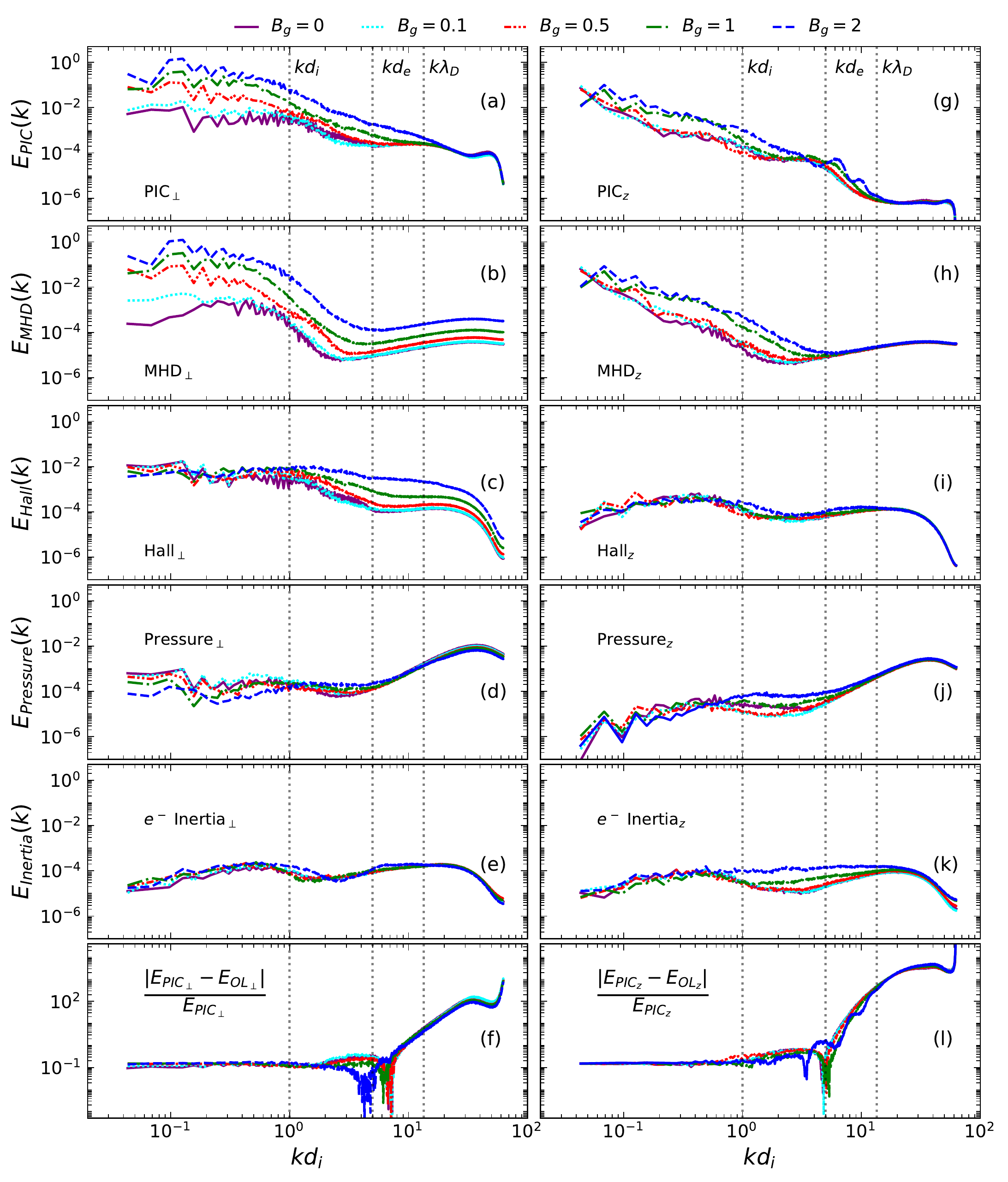}
\caption{\label{fig:ohmsspectra_para_perp}
Comparison of the perpendicular $(\perp)$ (left) and parallel $(z)$ (right) power spectra of the direct PIC output of the electric field (panels a, g), and different components of the generalized Ohm's law: MHD electric field (panels b, h), Hall electric field (panels c, l), pressure contribution of the electric field (panels d, j) and field due to electron inertia (panels e, k) for varying guide fields. The bottom panel (panels f, l)  compares the absolute value of the relative differences between the spectrum obtained summing the terms in the Ohm's law $(E_{OL})$ and direct PIC output $(E_{PIC})$. For the discussion (see text), we limit ourselves to the wavenumbers $k d_e \lesssim 1$.}%
\end{figure*}

The left panel of Fig. \ref{fig:ohmsspectra_g2_new} is the energy spectra for the different terms in Ohm's law (Eqn.~\ref{eqn:ohmslaw}) for the $B_g = 2$ simulation: (top) the perpendicular direction (i.e., $(x,y)$ plane); and (bottom) parallel (along $\hat{z}$). The time derivative in the electron inertia term is estimated by using successive time slices $1\,\omega_{ce}^{-1}$ apart. At the smallest $k$ in Fig.~\ref{fig:ohmsspectra_g2_new}, the spectra for both the perpendicular and parallel electric field are dominated by the MHD term of the electric field. The Hall electric field dominates the wavenumbers between the inverse of ion inertial length and the Debye length for $E_\perp$ perpendicular electric field, while  for $E_z$ the Hall term is significant in this range but not dominant. The electron inertia term is only significant for $E_z$ between the ion and electron inertial lengths. For the largest $k$ with $k \lambda_D > 1,$ the pressure term is the largest. 

Care must be taken when examining the spectra of different terms in Ohm's law. Notably, 
there is no guarantee that the spectra of individual terms of Ohm's law will sum to equal the spectrum of the electric field. From a theoretical standpoint, this equality requires that the different terms be uncorrelated or perpendicular to each other. This is evident from the right hand side of Fig.~\ref{fig:ohmsspectra_g2_new} which shows: (red) the spectra of the electric field from the direct output of the PIC simulation, (black) the sum of the spectra of the Ohm's law terms, and (gray) the spectra of the sum of Ohm's law terms. For $E_\perp$, all three curves are quite similar for $k d_e \lesssim 1$. At still smaller length scales, there is a large difference due to the pressure term. For $E_z$ on the other hand, the black curve begins to diverge at $k d_i \sim 1$ while the gray curve remains comparable for $k d_e \lesssim 1$. One major difference between Ohm's law for $E_\perp$ and $E_z$ is that for $E_z$ there are many terms that are comparable size for $k d_i \gtrsim 1.$ With many terms of comparable size, the effect of cross correlations of these terms is more important, which is a likely explanation for the divergence between the gray and black curve in this region. The results shown here agree with the one obtained from a turbulence simulation \cite{gonzalez2019turbulent} and magnetosheath observation \cite{stawarz2021comparative}.

For $k d_e \gtrsim 1,$ even the gray curve diverges. From an inspection of the left panels of Fig.~\ref{fig:ohmsspectra_g2_new} it is evident that the pressure term in Ohm's law is anomalously large in this region. Kinetic PIC simulations are known to generate significant numerical fluctuations at high $k$ \textcolor{red}~\cite{Birdsall85}. The calculation of the pressure term in Eqn.~\ref{eqn:ohmslaw} requires taking derivatives of pressure terms which exacerbates this noise at high $k.$ The direct output electric field is directly stepped forward in time in the PIC simulation using the current, and does not contain such a derivative. Note also, that the unphysical Ohm's law pressure term is primarily due to diagonal pressure terms for $E_\perp$ and off-diagonal terms for $E_z$. 

For the purposes of this paper, from now on we limit our discussion to regions where the PIC direct output is comparable to the Ohm's law sum, that is $k d_e \lesssim 1$. The top panel of Fig.~\ref{fig:ohmsspectra_para_perp} shows the spectrum of $E_\perp$ and $E_z$ calculated directly from the PIC output for all the different guide field cases. An increase in guide field increases the strength of both electric field spectra, although $E_\perp$ shows a larger increase than $E_z$. The shape of the spectra for $E_z$ are quite similar (right panels). In contrast, for $E_\perp$ the low guide field cases show a flat slope for $k\,d_i \lesssim 1$ while the higher guide field cases show a negative slope. Then, for higher $k$ all guide fields have a steepening to a strong negative slope.

Panels h-k show Ohm's law terms for $E_z$. For $k d_i < 1$, the MHD term dominates in all cases with the electric field much larger for larger guide fields. The increase in $E_z$ with guide field is consistent with the known increase in reconnection exhaust velocities with increasing guide field~\cite{haggerty2018reduction}. For $k d_i \sim 1,$ the MHD term continues to dominate for the large guide field cases, but in the smallest guide field cases the Hall term becomes comparable. This behavior is consistent with the reduction of the Larmor radius with guide field. For $k d_i >1,$ the Hall term gradually becomes dominant over the MHD term in all cases. The pressure term and the electron inertia term are never dominant in the range considered. This is because these terms become dominant at scales $d_e$ or smaller. The pressure term only becomes comparable to other terms in the highest guide field cases near $k d_e \sim 1$. The electron inertia term increases significantly with guide field between the wavenumbers $1<k d_i <5$. While the contribution from the electron inertia becomes comparable with the other terms for smaller guide field cases, it exceeds the contribution from other terms for the largest guide field case at around $k d_i > 2$.

Panels b-e show Ohm's law terms for $E_\perp$. As with $E_z$, the MHD terms are much larger for increasing guide field. However, for the smallest guide field cases, the Hall term is always larger than the MHD term; with very small guide field MHD physics generates very little $E_\perp$ while the Hall physics requires the generation of $E_\perp$. For $k d_i > 1$ the hall term gradually becomes larger than the MHD term even for the largest guide field cases. The pressure term is always negligible for $k d_e \lesssim 1$. The electron inertia term also becomes comparable around the electron scales $k d_e \sim 1$ but shows almost no dependence with the guide field.

The bottom panel shows the relative differences of the spectrum obtained directly from the PIC output with the spectrum of the sum of different terms in Ohm's law for both the parallel and perpendicular spectra. These values are in excellent agreement for $kd_i \lesssim 1$; for $k d_e > 1$ the relative difference increases drastically. Between $1 < k d_i < 5$, it exhibits a small increase because multiple Ohm's law terms become comparable in magnitude and presumably cross correlations are playing a significant role. The relative difference is further found to decrease with an increase in the guide field. Clearly, the choice of limiting the spectral discussion to wavenumbers $k d_e \lesssim 1$ is justified. 

\subsection{Energy transfer: von K\'arm\'an Howarth equation}
Using Kinetic PIC simulations, the behavior of the von-K\'arm\'an Howarth equation during anti-parallel reconnection was shown previously to have significant similarities to a generic decaying turbulence simulation \cite{adhikari2021magnetic}. The conclusion therefore was that in that antiparallel case, even laminar reconnection fundamentally involves an energy transfer to smaller scales (cascade). We now extend that analysis to the range of guide fields we have simulated. 

\begin{figure}
\includegraphics[scale=0.6]{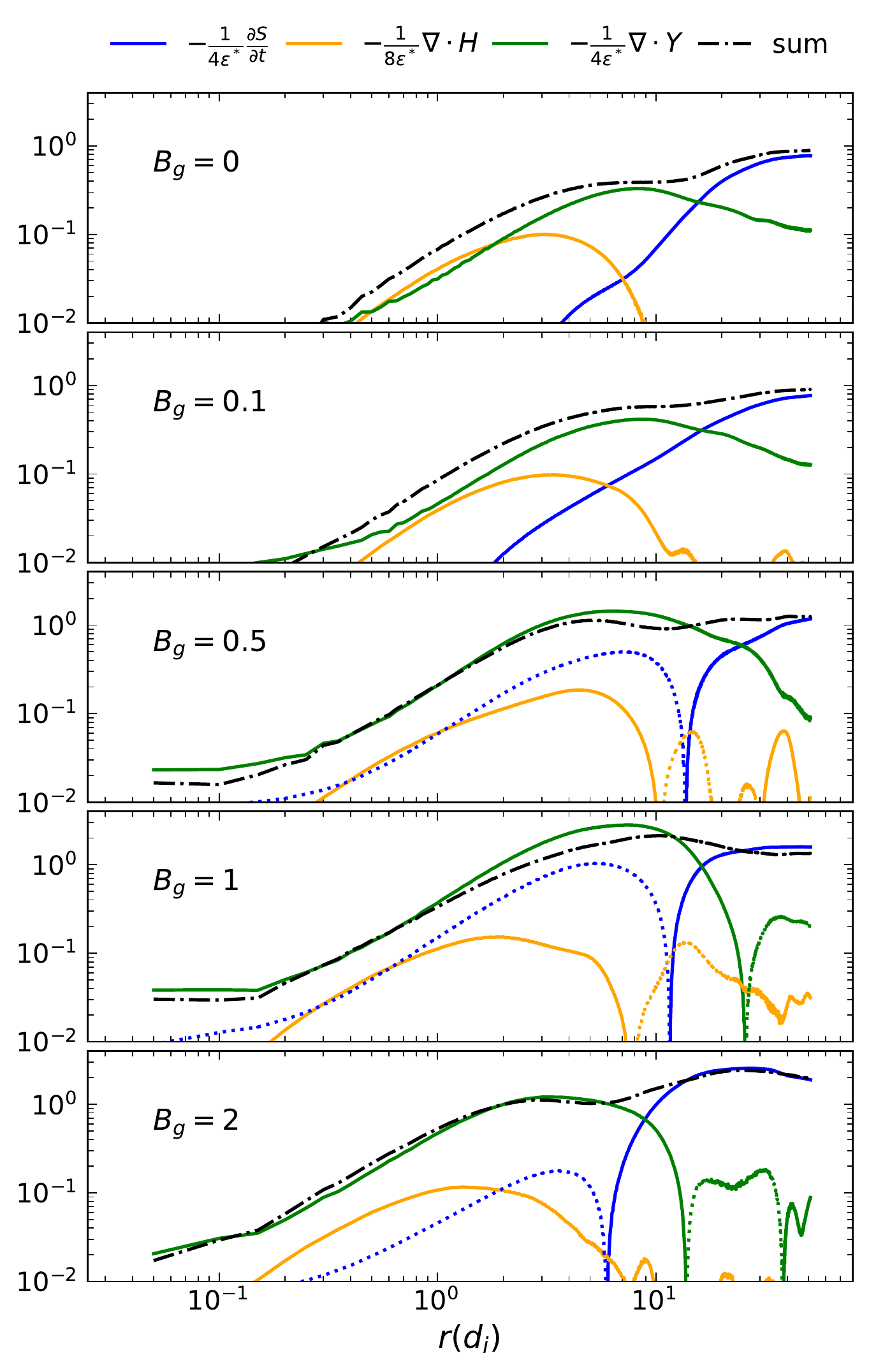}
\caption{\label{fig:thirdorderreconn_new} Individual terms of the MHD von K\'arm\'an Howarth equation and their sum, normalized to $\epsilon^*$ for all the simulations. A negative term is represented by a dotted curve of the same color.}
\end{figure}

In Fig.~\ref{fig:thirdorderreconn_new}, we plot the terms in the 
direction averaged form of the von-K\'arm\'an Howarth equation, as a function of lag magnitude. 
Note that these values are determined in the same way as done previously~\cite{adhikari2021magnetic}. The times analyzed all have the same amount of reconnected flux and are shown in Fig.~\ref{fig:jzreconn}. For each of these times, the time rate of change of the sum of 
magnetic fluctaution energy and ion flow energy is calculated and denoted $\epsilon^*$. First, the general behavior of the different terms remains quite similar when the 
guide field is varied. The $\partial S/\partial t$ terms dominates at largest scales, the  MHD transfer term ($-\nabla_l \cdot \textbf{Y} $) dominates at intermediate scales, while the Hall transfer term ($-\nabla_l \cdot \textbf{H}$) reaches it peak value at scales near the inertial length. The decrease at smallest scales of the sum of the terms is due to the importance of dissipation at these scales. The kinetic systems do not have a closed form lag dependent dissipation function in terms of increments, even if the total dissipation is well accounted for by the pressure work
\cite{yang2022pressure}.
Since we are not accounting for 
scale dependent dissipated energy, 
the sum of other terms falls short of the total $\epsilon^*$. As higher order models, such as compressible Hall MHD or compressible two fluid MHD, are considered, the regime of validity can be pushed to smaller scales~\cite{banerjee2020scale, hellinger2021spectral}. Another empirical approach is to introduce pressure strain interactions~\cite{yang2022pressure} 
as an approximation~\cite{hellinger2022ion}.

Second, the crossover point of dominance between the $\partial S/\partial t$ terms and the MHD transfer does not show a clear pattern with changing guide field. The two largest guide field cases continuously generate secondary islands which create complexity around the scale of about $10$ ion inertial lengths. Third, there is a clear trend that the MHD transfer term remains large to smaller lags with increasing guide field. This trend is consistent with the reduction in particle Larmor radius with increasing guide field, which makes MHD physics dominant at smaller scales. Fourth, however, the importance of the Hall transfer term is reduced with larger guide field. Apparently, the scale at which dissipation is occurring is not necessarily shrinking as the guide field increases. This effect leads to a "squeezing" in lag of the Hall transfer, and a reduction in its importance for the largest guide field cases. 

While the energy transfer has many similarities between the different guide field cases, the question remains if the anisotropy of the magnetic spectrum will behave the same. In antiparallel reconnection it was shown that the anisotropy of the magnetic spectrum steadily decreased with time, moving energy from $k_y$ to $k_x$~\cite{adhikari2020reconnection}.

\begin{figure*}
\includegraphics[scale=0.7]{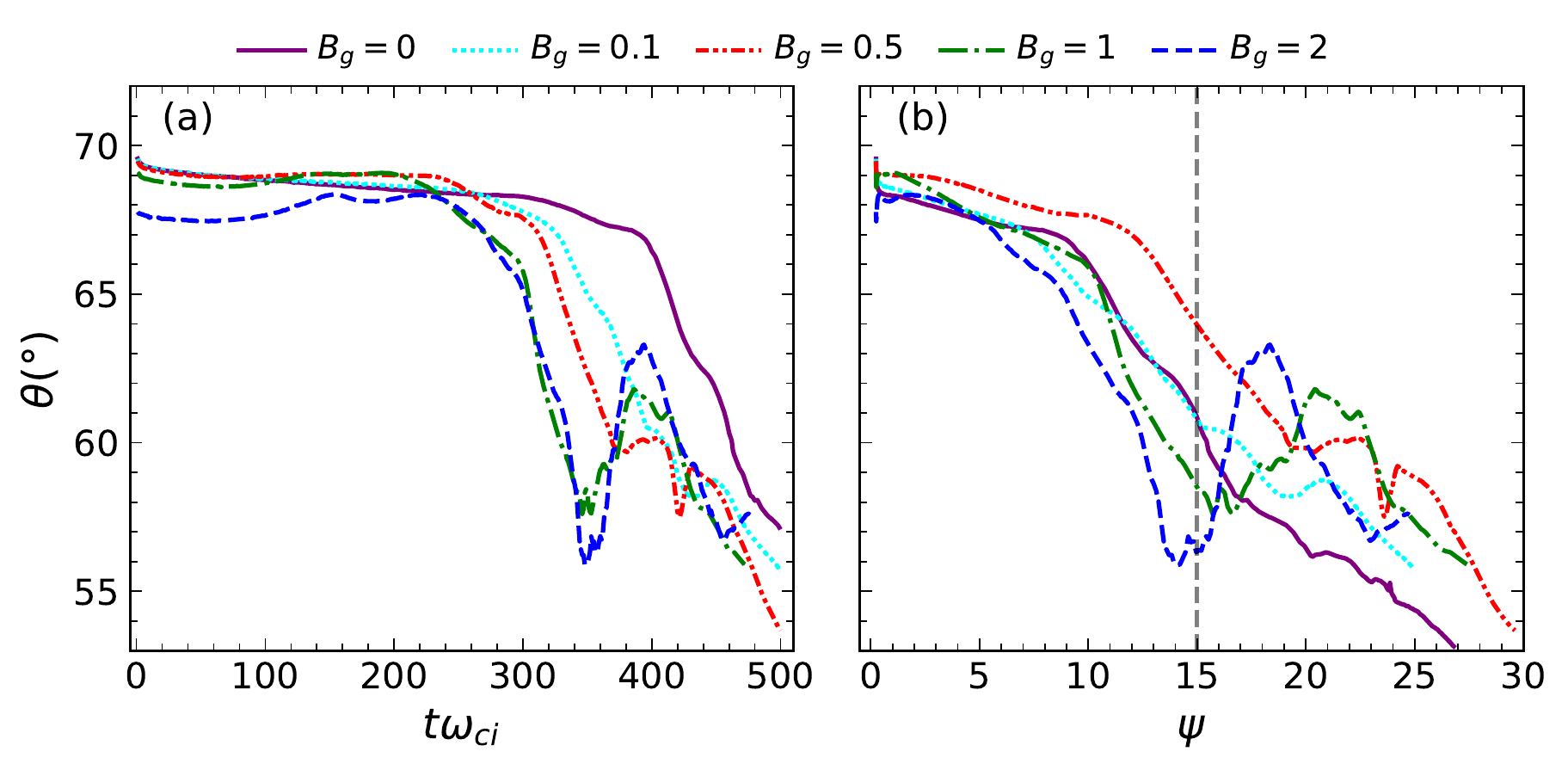}
\caption{\label{fig:shebalin_time} (a) Time evolution of the Shebalin angle, and (b) Shebalin angle versus the reconnected flux $(\psi)$ for all the simulations. The vertical dashed line at $\psi = 15$ represents the flux chosen for the analysis.}
\end{figure*}

The anisotropy was measured in terms of the Shebalin angle~\cite{shebalin1983anisotropy} corresponding to the magnetic spectrum, defined as:
\begin{equation}
    tan^2\theta_{B} =\frac{\sum \limits_{k_{x},k_{y}} k_{y}^2|E_{B}(\vec{k},t)|}{\sum \limits_{k_{x},k_{y}} k_{x}^2|E_{B}(\vec{k},t)|},
\end{equation}
where $k_{x}$ and $k_{y}$ correspond to the wavenumber along $x$ and $y$ axes respectively while $E_{B}$ is the 2D magnetic energy spectrum. For reference, a Shebalin angle of $45^{\circ}$ represents isotropy. Figure.~\ref{fig:shebalin_time}a shows the time evolution of the Shebalin angle across all the simulations. Initially, since the magnetic energy is dominated by the wavenumbers along y-axis $k_y$, the Shebalin angle is close to $90^{\circ}$. However, with the onset of reconnection the energy along $k_y$ is redistributed along $k_x$ as suggested by the decrease of the Shebalin angle. The larger guide field runs have earlier reconnection onset and therefore an earlier reduction of the Shebalin angle. One peculiar property observed in the higher guide fields runs is the transient increase of the Shebalin angle. This behavior occurs because secondary islands readily form in the higher guide field simulations. Both the $B_g = 1$ and $2$ cases generated large secondary islands that reached scale sizes of about $20$ ion inertial lengths before they were absorbed into the main magnetic islands. These local islands transfer significant energy from $k_y$ to $k_x$, some of which transfers back when they are absorbed, leading to the transient increase in the Shebalin angle. 

Unlike the time evolution of energies in Fig.~\ref{fig:energyrecon}, plotting the Shebalin angle versus reconnected flux in Fig.~\ref{fig:shebalin_time}b doesn't simplify the figure substantially. At $\psi \approx 10,$ for example, $B_g = 0.5$ has the largest angle while $B_g = 2$ has the smallest. It seems clear that secondary island generation significantly complicates the system anisotropy. 

\section{\label{sec:conclusions}Conclusions}
In this paper, using $2.5$D kinetic PIC simulations, we have studied the effect of a guide field on the spectral and energy transfer properties of magnetic reconnection. The key result is that the energy transfer properties of reconnection show the same qualitative behavior independent of guide field, which implies that magnetic reconnection fundamentally involves 
energy transfer, or in vernacular, 
an energy cascade,
that is in a number of aspects much the same as what is attributed commonly to turbulence. 

The Kolmogorov-like spectral index observed previously in the magnetic energy spectrum of an anti-parallel reconnection is persistent even with the addition of a guide field. While the slope of the magnetic energy spectrum stays roughly the same in the inertial range $(k d_i \lesssim 1)$, the kinetic range slope becomes steeper for lower guide field cases. Conversely, the electric field energy spectrum displays significant changes with the guide field. For wavenumbers in the range $0.1 \lesssim k d_i \lesssim 1$, the slope of the electric field spectrum decreases from about zero to almost a $-5/3$ for the largest guide field case $B_g=2$. 

The variation in the electric field spectra with guide field is further quantified by decomposing the electric field into various terms of the generalized Ohm's law. For the $B_g=2$ case, we find that the MHD electric field dominates the perpendicular and parallel electric field at the smallest $k$. The Hall electric field, however, dominates the wavenumbers between $k d_i$ and $k \lambda_D$ for $E_\perp$. For $E_z$, it becomes significant and comparable to the pressure and electron inertia contributions. The electric field due to the pressure and electron inertia is significant only in the parallel direction between $k d_i$ and $k d_e$. It is important to note that the largest wavenumbers are often polluted with the finite particle in the grid effect. Therefore, one must be careful while examining the spectra of individual terms in Ohm's law. As a result, we compare (a) the spectrum of the electric field directly obtained from the PIC simulation, (b) the sum of the spectrum of individual terms in the Ohm's law, and (c) the spectrum of the sum of different terms in the Ohm's law and restrict ourselves to wavenumbers where the spectrum of the PIC output is comparable to the spectrum of sum of terms in Ohm's law. Further, one must also use a high cadence data to study the electron inertia terms in Ohm's law such that the time scales at which the current density change are properly addressed. 

The variation of the electric field spectrum with guide field is primarily due to $E_\perp$, which shows almost a flat slope for $B_g=0$ and an increasing negative slope for higher guide fields. At smaller wavenumbers $(k d_i <1)$, the MHD electric field dominates the parallel and perpendicular spectrum for all the guide field cases. For $E_z$, the Hall field plays a significant role for the $B_g=0$ case at $k d_i \approx 1$, while MHD contribution still dominates for the large $B_g$. Whereas, for $E_\perp$, the Hall term consistently exceeds the MHD term in the smaller wavenumbers. 
The parallel electric field due to the pressure and electron inertia is insignificant for all but $B_g=2$ case, where it exceeds the contributions from all other terms in the Ohm's law at scales around $k d_i >2$. In the perpendicular field, however, the pressure term is negligible. The electron inertia contribution becomes comparable in the electron scales but does not vary much with the guide field. These results 
of examining the spectra of the contributions to the generalized Ohm's law are consistent with the conclusions obtained from a kinetic PIC simulation of turbulence~\cite{gonzalez2019turbulent,adhikari2021beta,adhikari2022reconnection}, as well as MMS observations~\cite{stawarz2021comparative}.

Finally, we also explore the effect of an external field on the von-K\'arm\'an Howarth equation. The energy transfer characterstics in guide field reconnection are qualitatively the same as what is observed in anti-parallel reconnection as well as fully developed MHD and kinetic turbulence. The largest scales are dominated by the $\partial S/\partial t$ term; the intermediate scales are dominated by the $-\nabla_l \cdot \textbf{Y}$, and at the smaller scales, the Hall term $-\nabla_l \cdot \textbf{H}$ becomes significant. The total sum of these energy transfer terms, when normalized to the rate of change of magnetic and ion-flow energy $\epsilon^*$, is close to unity at the larger scales. At the smallest scale, however, the sum falls significantly because of the absence of a proper description of dissipation in collisionless plasma. While pressure interaction such as $Pi-D$~\cite{yang2017energya} is considered a possible candidate to describe dissipation in collisionless plasmas, it involves scale-filtering techniques~\cite{yang2022pressure} which is out of the scope of this paper and is left for a subsequent study. 

Comparing the energy transfer behavior in different guide field simulations can be a delicate matter. For example, the degree of anisotropy in the systems under comparison might vary. Because of an earlier onset of reconnection, the higher guide field runs seem to have an 
isotropization of energy in $k_x$ $k_y$ space (see Fig.~\ref{fig:shebalin_time}). The Shebalin angle for the higher guide field runs also displays a transient increase after reaching the first minimum. This feature is due to the formation of secondary islands which lead to a temporary transfer of energy back to $k_x$.

 We have investigated spectra and 
 energy transfer and the effect of  the guide field on these, 
 within the context of the 
 incompressible Hall MHD von-K\'arm\'an Howarth equation. 
 Although this equation remains valid for non-isotropic systems \cite{hellinger2018karman, ferrand2019exact,adhikari2021magnetic,wang2022strategies},
 it does not account for compressive effects on spectral transfer.
 We are aware of the recent developments of the exact laws of energy transfer in compressible systems \cite{banerjee2013exact,andres2018energy,banerjee2020scale,hellinger2021spectral}. But we consider implementation of these more complex formulations,
 which also adopt additional simplifying assumptions, to be beyond the scope of the present paper. 
 We leave examination of compressive channels of transfer in kinetic plasma reconnection to future study. 
The 
main findings here are 
that spectral and energy transfer in the nonlinear phase of laminar collsionless reconnection,
with a range of applied guide fields,
bears considerable fundamental resemblance to properties of incompressive MHD turbulence. 
This recognition
may guide future investigations 
that seek to unify 
perspective on the physics of these two important plasma processes.

\clearpage
\begin{acknowledgments}
The authors would like to thank Paul Cassak for the fruitful discussions. We also acknowledge the high-performance computing support from Cheyenne~\cite{Cheyenne18} provided by NCAR’s Computational and Information Systems Laboratory, sponsored by the NSF. This research also used NERSC resources, a U.S. DOE Office of Science User Facility operated under Contract No. DE-AC02-05CH11231. S.A., and M.A.S. acknowledge support from NASA LWS 80NSSC20K0198. W.~H.~M is supported by NSF DOE grant 
AGS 2108834 at the University of Delaware, by the IMAP project (Princeton subcontract SUB0000317)) and by the NASA LWS program FST grant to New Mexico Consortium (subcontract 655-001 to Delaware).
J.~E.~S. is supported by Royal Society University Research Fellowship URF\textbackslash R1\textbackslash 201286, and J.~P.~E. is supported by UKRI/STFC grant {ST/W001071/1}. 
\end{acknowledgments}

\section*{Data Availability Statement}

The datasets are available from the corresponding author [SA] upon reasonable request.

\bibliographystyle{abbrv}
\bibliography{aipsamp}

\end{document}